\documentclass[%
reprint,
superscriptaddress,
 amsmath,amssymb,
 aps,
 prl,
]{revtex4-2}

\usepackage{graphicx, subcaption}% Include figure files
\usepackage{dcolumn}% Align table columns on decimal point
\usepackage{bm}% bold math
\usepackage{gensymb} %degree symbol, added by Martin
\usepackage{booktabs} %table setup, added by Martin. Not sure if this is "ok" for PRL
\usepackage{siunitx} %typesetting units
\usepackage[dvipsnames]{xcolor} %adding comments in color 
\usepackage[version=4]{mhchem} %\ce command for chemical formula

\DeclareCaptionJustification{justified}{\leftskip=0pt \rightskip=0pt \parfillskip=0pt plus 1fil}
\begin{document}

%\preprint{APS/123-QED}

% --------------------------------------------------------------------------- %
% Title
% --------------------------------------------------------------------------- %
\title{Observation of the Boson Peak in a 2D Material}% Force line breaks with \\

\author{M. T{\o}mterud}
\affiliation{Department of Physics, NTNU--Norwegian University of Science and Technology, 7491 Trondheim, Norway}
\affiliation{Department of Physics and Technology, University of Bergen, All\'egaten 55, 5007 Bergen, Norway}

\author{S. D. Eder}
\affiliation{Department of Physics and Technology, University of Bergen, All\'egaten 55, 5007 Bergen, Norway}

\author{C. B\"uchner}
\affiliation{Fritz-Haber-Institut der Max-Planck-Gesellschaft, Faradayweg 4--6, 14195 Berlin, Germany}

\author{M. Heyde}
\affiliation{Fritz-Haber-Institut der Max-Planck-Gesellschaft, Faradayweg 4--6, 14195 Berlin, Germany}

\author{H.-J. Freund}
\affiliation{Fritz-Haber-Institut der Max-Planck-Gesellschaft, Faradayweg 4--6, 14195 Berlin, Germany}

\author{I. Simonsen}
\affiliation{Department of Physics, NTNU--Norwegian University of Science and Technology, 7491 Trondheim, Norway}

\author{J. R. Manson}
\affiliation{Department of Physics and Astronomy, Clemson University, Clemson, South Carolina 29634, U.S.A.}
\affiliation{Donostia International Physics Center (DIPC), Paseo Manual de Lardizabal, 4, 20018 Donostia-San Sebasti\'an, Spain}

\author{B. Holst}
\affiliation{Department of Physics and Technology, University of Bergen, All\'egaten 55, 5007 Bergen, Norway}

\date{\today}% It is always \today, today,

% --------------------------------------------------------------------------- %
% Abstract
% --------------------------------------------------------------------------- %
\begin{abstract}
The boson peak is an excess in the phonon density of states relative to the Debye Model, which occurs at frequencies below the Debye limit. It is present in most amorphous materials and, as was recently shown, can sometimes be found also in crystals. Here we present first experimental evidence of the boson peak in a 2D  material, namely 2D silica (\ce{SiO2}). The measurements were obtained by helium atom scattering. A dispersionless boson peak is seen at $6 \:\pm 0.5$~meV (1.5$ \:\pm$ 0.15 THz) and $-6 \: \pm 1.5$~meV (-1.5 \:$\pm$ 0.4 THz), with reasonable evidence for a double excitation at  $ \pm 12 \: \pm 2.5$ meV (2.9 $\pm$ 0.6 THz). 
\end{abstract}

%\keywords{randomly rough surface, rough surface scattering, inverse scattering problem, surface-height autocorrelation function, Kirchhoff approximation}
%\keywords{Suggested keywords} % Use showkeys class option if keyword
                               % display desired

\maketitle

%--------------------------------------------------------------------
%   MAIN TEXT
%--------------------------------------------------------------------

The Debye model predicts that below the so called Debye limit the phonon density of states of a bulk material should be proportional to the phonon frequency squared~\cite{Debye1912}. Experimentally, it has been shown that many materials exhibit a characteristic deviation from this: a broad, dispersionless excess of states in the phonon density of states relative to the expected frequency squared dependence. 
Usually it is seen as a broad peak in the reduced vibrational density of states, which is the density of states divided by the squared frequency.
This broad peak is commonly referred to as the \textit{boson peak}~(BP)~\cite{MALINOVSKY1986757,PhysRevB.34.5665,PhysRevLett.92.245508,Elliott2001}. The BP excess in the phonon density of states  
implies a corresponding excess in the real heat capacity of the material~\cite{Buchenau2001,Pohl2001}. That is to say, in the BP phonon frequency range, the heat capacity is larger than predicted from the Debye model.  

Experimentally, the BP has been observed in the bulk of numerous disordered or glassy materials, using a wide range of measurement techniques: Raman spectroscopy~\cite{MALINOVSKY1986757,PhysRevB.94.224204}, optical spectroscopy~\cite{SCHROEDER1997342} including far-infrared spectroscopy~\cite{Hutt_Far_Infrared, PhysRevB.50.9569}, X-rays~\cite{PhysRevLett.77.3835}, neutron scattering~\cite{neutron:doi:10.1080/13642819808204986,Buchenau_2019}, as well as  thermal techniques~\cite{Tomoshige_2019},  including indirect verification through measurements of the temperature dependence of the heat capacity~\cite{1981amorphoussolids,PhysRevB.4.2029, PhysRevB.29.4778}.  In 2007 the BP was observed for the first time as a surface phenomenon through measurements on vitreous silica using helium atom scattering (HAS)~\cite{PhysRevLett.99.035503, PhysRevB.78.045427, PhysRevLett.100.135504, Steurer_2008}. HAS has unique surface sensitivity, with no penetration into the bulk and is therefore particularly suited for this type of measurement~\cite{Holst2021, Holst2013}. The surface BP was found at an energy of about $4$~meV roughly corresponding to 1~THz, in good agreement with theoretical predictions~\cite{Wang2003}. In a related publication, the surface BP was found to be heavily temperature dependent, blueshifting with increasing temperature even faster than observed for the BP in the bulk~\cite{PhysRevLett.100.135504}. Note that in 2D the Debye model predicts the phonon density of states to be linearly proportional to the phonon frequency.

Theoretically, one has long assumed some form of disorder in a material to be a prerequisite  for the existence of the BP. 
Early approaches used soft potentials and double-well potentials introduced into the Hamiltonian and these are often called soft potential models \cite{Karpov,Buchenau-2,Buchenau-3,Parshin,Buchenau}.
Another example is the heterogenous elasticity theory proposed by Schirmacher, Ruocco, and  Scopigno~\cite{Schirmacher2006, Schirmacher2007, Schirmacher2015}. Other examples include more recent theories focusing on local inversion symmetry breaking~\cite{Krausser2017, PhysRevB.93.094204}. A 2014 observation of the BP in crystalline $\alpha$-quartz~\cite{PhysRevLett.112.025502} was therefore unexplained for several years, until in 2019  Baggioli and Zaccone~\cite{PhysRevLett.122.145501} proposed a universal model for the BP in both crystalline and vitreous materials. 
This model focuses on anharmonicity as the root cause for the occurence of the BP. 
It has also been suggested that in glasses acoustic van Hove singularities can be responsible for the Boson peak~\cite{Chum-2011}.  A recent paper shows that the BP is related to phonon attenuation due to quasilocalized modes~\cite{Ciamarra}.

%on damping of phonon vibrations as the root cause for the occurrence of the BP. The damping in perfectly ordered crystals emerges from including an anharmonic term in the Hamiltonian of the system~\cite{BottgerHarald1983Pott} and computing the corresponding phonon Green's function~\cite{LoveseyS.W1980Cmp:}, from which the vibrational density of states can be extracted. For vitreous materials, so the model, the damping can be attributed to the disorder of the material. The competition between propagation at low frequencies and diffusive damping at higher frequencies is cited as the mechanism responsible for generating the BP. 
 
The BP has been predicted as a $2$D phenomenon by 2D random matrix models~\cite{Conyuh_2017, Conyuh2020, Raikov2020} and has been observed in a 2D macroscopic model system consisting of photo-elastic disks~\cite{Zhang2017,PhysRevB.98.174207}. It was recently claimed to have been observed in ultrathin alumina layers on oxidized aluminum nanoparticles~\cite{PhysRevResearch.2.023320}, but has up till now not been experimentally observed in a $2$D material. 
 
Here we present the first experimental measurements of the BP in an atomically thin 2D material, namely a bilayer of amorphous silica supported on Ru(0001).
This film system is a transferable wide band gap 2D material.
The 2D silica on Ru(0001) was chosen as the first system to investigate because it has been shown that the silica is very weakly bound to its supporting substrate: it can be peeled off~\cite{BUCHNER2017341} and in Ref.~\onlinecite{PhysRevLett.120.226101} it is shown that mechanical properties of free 2D silica can be extracted from the phonon frequency spectrum.
Furthermore, as discussed in Ref.~\onlinecite{PhysRevLett.120.226101}, the phonon spectrum exhibits no prominent features that can be directly related to the underlying substrate, for example no Rayleigh mode is visible. However, it should be noted that there are cases where the underlying substrate has been observed to have an influence on the low energy phonon spectrum for a 2D material. In particular, the electron-phonon coupling constant $\lambda$ has been observed to increase with decreasing binding energy for graphene~\cite{Benedek2021}, thus we cannot claim that the phonon spectrum we measure will correspond exactly to a measurement on free standing 2D silica.
The 2D silica sample was prepared in UHV on a Ru(0001) substrate at the Fritz-Haber-Institut in Berlin.
We used the same sample that is described in Ref.~\onlinecite{PhysRevLett.120.226101}.
According to scanning tunneling microscopy (STM) and low energy electron diffraction (LEED) measurements presented in Ref.~\onlinecite{PhysRevLett.120.226101} the sample exposes crystalline and vitreous silica film patches.

After preparation, the sample was removed from the UHV chamber in Berlin and transported to Bergen. During transport, the sample was exposed to ambient conditions for more than 20 hours. 
Upon arrival from Berlin, the sample was installed in the argon vented sample chamber which was then pumped down. The background pressure was around 1$\times$10$^{-9}$~mbar. A signal could be obtained from the sample without any initial cleaning, however, to ensure maximum intensity before measurements were done, the sample was heated to 675~K for one hour in an oxygen atmosphere ($p_{\ce{O2}}=2.2\times10^{-6}$~mbar). This improved the measured signal. A slight decline in reflected signal could be observed over a period of days. For this reason, the cleaning process was repeated every day before measurements. This restored the original reflectivity.
The HAS experiments were carried out in MAGIE, the molecular beam apparatus at the University of Bergen \cite{Apfolter,Eder}. 
The neutral helium beam was created by a free-jet expansion from a source reservoir through a 10$\pm$0.5~\si{\micro \meter} diameter nozzle. The central part of the beam was selected by a skimmer, 410$\pm$2~\si{\micro \meter} in diameter, placed 18$\pm$0.5~\si{\micro \meter} in front of the nozzle. All experiments presented here were carried out on a room temperature sample ($T=296\pm$1~K). The source-detector scattering angle was held constant at $90^{\circ}$. The diffraction scans were measured with a room temperature beam corresponding to a beam energy of $E_0$=64~meV, while for time-of-flight (TOF) measurements the beam was cooled to an energy of around $E_0$=29~meV. The stagnation pressure in the source reservoir was $p_0$=81~bar for all TOF measurements. The incident beam size was around 4 mm in diameter at the sample position. 
TOF measurements were performed with a pseudorandom chopper~\cite{Koleske}. The measurements presented were conducted as six independent series, varying the incident beam angle (polar angle) $\theta_i$ at five different azimuthal angles, $\phi$. The azimuthal angle was varied in order to ensure that the effects measured were truly anisotropic. 
However, no crystalline diffraction pattern was observed with HAS, see measurements also presented in Ref.~\onlinecite{PhysRevLett.120.226101},
which suggests that the crystalline fraction is small
and that the sample is primarily amorphous.
The thickness of an atomically thin layer is difficult to define, a problem well-known in the study of single-layer graphene where it called the Yakobson paradox, however, based on density functional theory calculations we have estimated the bilayer thickness to be about 6.1 \AA~\cite{PhysRevLett.120.226101}.
In total 156 TOF spectra were obtained.

\begin{figure}[!tbh]
%\begin{center}
    \centering
\includegraphics[width = \columnwidth]{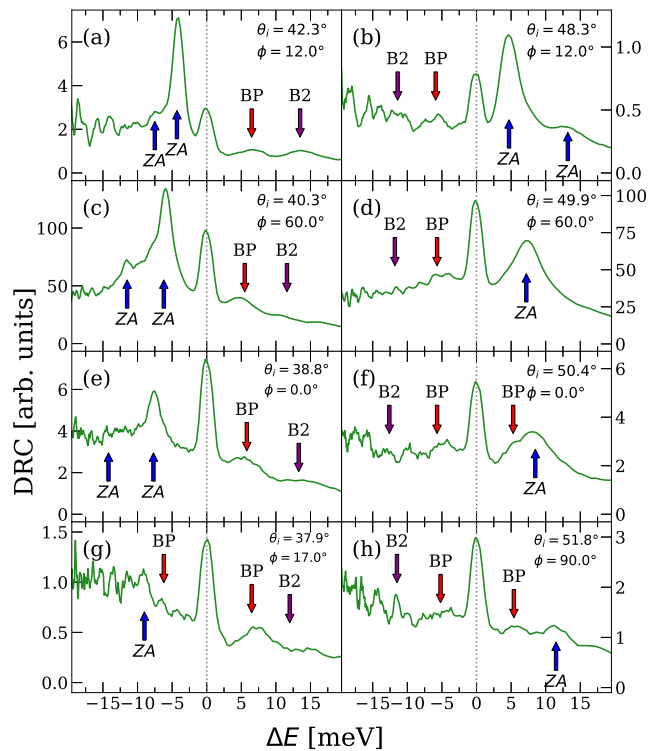}
%\end{center}
\caption{Several examples of helium atom scattering TOF spectra, converted to energy transfer $\Delta E$, for 2D silica supported on Ru(0001).  The incident polar angle relative to the sample normal is $\theta_i$ and $\phi$ is the azimuthal angle of the sample relative to the principal axis of the Ru(0001). The upward-pointing blue arrows indicate the shear vertical (ZA) mode, the downward-pointing red arrows indicate the BP and the downward-pointing purple arrows indicate the double excitation of the BP, labelled B2.}
    \label{fig:fig1}
\end{figure}

Figure~\ref{fig:fig1} shows examples of TOF spectra converted to energy transfer $\Delta E$.
These examples were chosen so as to cover the complete range of incident polar angles explored.
The peak at $\Delta E=0$ corresponds to diffuse elastic scattering, where there is no energy transfer with the surface. Negative values of $\Delta E$ correspond to phonon creation, positive values to phonon annihilation. Phonon mode peaks are indicated with arrows. The upward-pointing blue arrows indicate the shear vertical
mode (ZA) which was analyzed in detail in~Ref.~\onlinecite{PhysRevLett.120.226101} where it was used to obtain the bending rigidity of 2D silica. The downward-pointing red arrows indicate 
the BP mode and the downward-pointing purple arrows marked as B2 indicate the double excitation of the BP.

%----------------------------------------------------------------------------
%-----------------------------------------------------------------------------

The TOF spectra, once they are converted to energy transfer, are 
known as the
differential reflection coefficient (DRC) $ {dR}/{d\Omega dE}$: the number of helium atoms $dR$ scattered within a small solid angle $d\Omega$ and small energy interval $dE$.
To a good approximation the DRC is given by~\cite{Manson-91,PhysRevB.78.045427}
\begin{eqnarray} \label{eq1}
\frac{dR}{d \Omega dE}  \propto
\frac{|{\bf k}_f|}{k_{iz}} 
e^{-2W(T)}~ 
%\exp{[-2W(T)]} 
|\tau_{fi}|^2
~
%\nonumber \\ \times ~~
\Delta {\bf k} \cdot
\underline{\underline{{\bf \rho}}} \cdot \Delta {\bf k}
~
|n(\omega)|,\;\;\;
\end{eqnarray}
 {where ${\bf k}_f$ is the wave vector of the helium atom after the scattering event, $k_{iz}$ is the $z$-component of the initial wave vector of the helium atom,  the scattering vector is $ \Delta{\bf  k} = {\bf k}_f - {\bf k}_i$,  $\tau_{fi}$ is the form factor of the interaction potential between the helium atom and the 2D-silica, and $n(\omega)$  {is} the Bose-Einstein function.  The dyadic $\underline{\underline{{\bf \rho}}}$ is the surface phonon spectral density.
The Bose-Einstein function is given by $n(\omega) = \left( \exp[\beta \hbar \omega] -1 \right)^{-1}$, where the energy transfer of the helium atom is given by $\Delta E = \hbar \omega$, and $\beta = (k_B T)^{-1}$ with $k_B$ being the Boltzmann constant and $T$ the substrate temperature.
The expression $\exp[-2W(T)]$ is the  Debye-Waller factor.
}

 {
For experimental conditions such as the case here in which the incident and final polar scattering angles are near the specular position, the perpendicular scattering vector $\Delta k_{z} = |k_{fz}|+|k_{iz}|$ is much larger than the parallel wave vector transfer.  Thus, to a very good approximation the dominant contribution to the phonon spectral density is given by
}
\begin{equation}  
        \rho_{zz} \propto \frac{\left[ {dR^{expt}}/{ d \Omega dE} \right]|k_{iz}|}{|{\bf k}_f| ~|\tau_{fi}|^2 \Delta k_z^2
       ~ e^{-2W(T)} ~ |n(\omega)|  },
    \label{eq:rho_zz_methodology}
\end{equation}
 {where ${dR}^{expt}/{d\Omega dE}$ is the experimentally measured differential reflection coefficient
and $\Delta k_z$ is the perpendicular component of the scattering vector $\Delta {\bf k}$.
In the 2D Debye phonon model $\rho_{zz}$ is linear in $\omega$.}

 The form factor $\tau_{fi}$ is approximated using the well-known Morse potential
matrix elements, i.e., the distorted wave matrix elements of the Morse potential taken with respect to its own eigenstates~\cite[Ch. 8]{1976goodmanwachman}.
The argument of the Debye-Waller factor, $2W(T)$, within the harmonic lattice approximation and for sufficiently large temperatures is linear in $T$, and is usually determined by measuring the temperature dependence of the specular diffraction peak.  In the present set of experiments, the temperature dependence of the specular peak intensity was not measured, so the argument was taken to be represented by $2W =(\Delta {\bf  k}^2 C + D) T$ where the coefficients $C$ and $D$ were taken as  fitting parameters.  
The coefficient $D$ is the so-called Beeby correction, in which the attractive well depth of the interaction potential is added to the energy of normal motion  of the He atom~\cite{Beeby}.
As a check on this procedure previous data from which the BP was determined at a silica surface~\cite{PhysRevLett.99.035503} were re-examined using this method, and nearly identical results were found for both the Debye-Waller factors and the surface BP intensities.
%Thus, this fitting procedure turns out to be a novel method for determining the Debye-Waller factor and is discussed in more detail in  Ref.~\onlinecite{Martin-21}.}

\begin{figure}[!bth]
    \centering
    \includegraphics[width = \columnwidth]{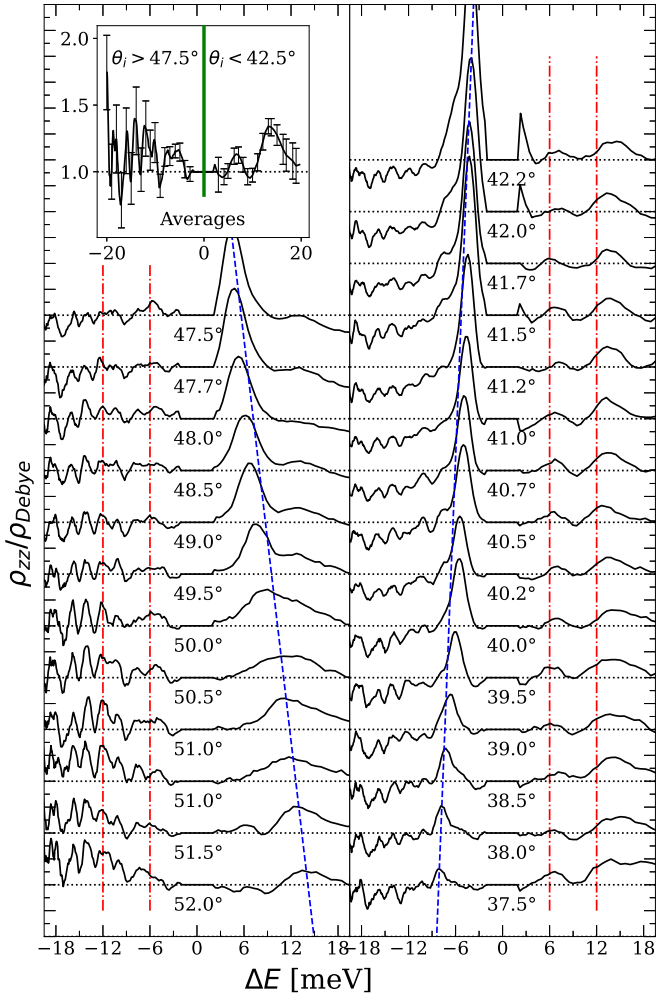}
    \caption{The phonon spectral density normalized to the Debye model approximation as a function of $\Delta E$.  
   The individual spectra shown are a selection taken from those points  exhibited in Fig.~\ref{fig:fig3} (a) covering incident polar angles larger than specular (left side) and smaller than specular (right side).  
    The region in the neighborhood of the strong elastic peak at $\Delta E = 0$ has been subtracted away.
    Incident angles very close to specular at $\theta_i = 45^\circ$ are not shown because the very intense ZA mode hides the BP.
    The blue dashed curve marks the dispersive ZA mode and the vertical red dashed lines mark the BP and B2 positions, which always remain at the same energy.  The insert shows the averages of the BP and B2 with error bars of one standard deviation.}
    \label{fig:fig2}
\end{figure}

\begin{figure}[!tbh]
    \centering
    \includegraphics[width = \columnwidth]{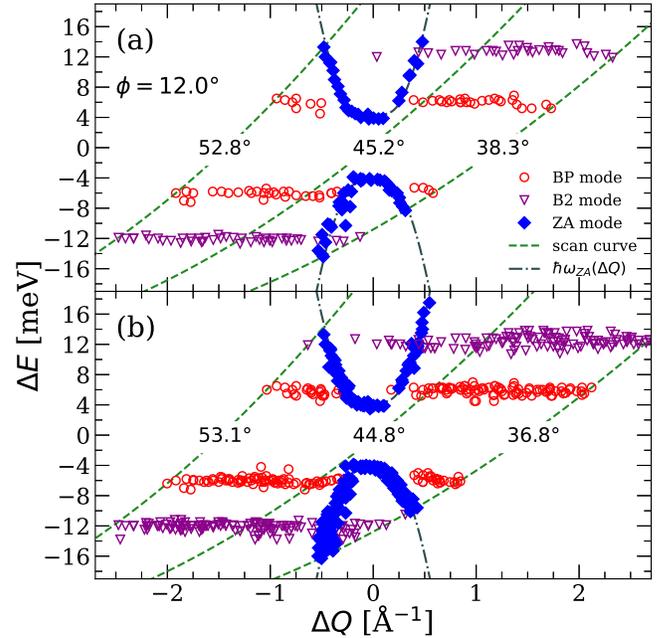}
    \caption{Phonon dispersion curves showing energy transfer $\Delta E$ as a function of parallel momentum transfer $\Delta Q$. (a) Dispersion curve for all polar angles $\theta_i$ measured at the single azimuthal angle $\phi = 12^\circ$.  The BP is shown as red circle data points, the ZA mode is blue diamonds and the double BP excitation is indicated by inverted purple triangles.  The dotted green lines show the scan curves for three selected $\theta_i$ as indicated.  (b) Dispersion curves including all polar and azimuthal angles measured.}
    \label{fig:fig3}
\end{figure}

The next step is to compare the obtained phonon spectral densities with the linear frequency dependence predicted in 2D by the Debye model. This is done in Fig.~\ref{fig:fig2} 
which shows a large number of reduced phonon spectral density spectra taken from
points on the dispersion curve shown in Fig.~\ref{fig:fig3} (a) which is for an azimuth of $12^\circ$ and incident polar angles ranging from $\theta_i = 37.5^\circ$ to $52^\circ$.  In each spectrum the region close to the large elastic peak at $\Delta E = 0$ has been subtracted out. The ZA mode is marked by the dashed blue curve and disperses to higher energy values as the incident angle moves further away from the specular position, which corresponds to larger values of $|\Delta Q|$.  The BP appears at the same energy of about 6 meV (corresponding to mode annihilation) on the right side of Fig.~\ref{fig:fig2} and about -6 meV on the left (creation).  It is denoted by a vertical red dashed line and has a FWHM of about 3.5 meV.  
As discussed below in connection with Fig.~\ref{fig:fig3} a He atom experiment measures only those inelastic features that cross its scan curve, and this explains why, in the right side of Fig.~\ref{fig:fig2}, the BP appears only at positive energy transfer and the ZA mode appears only at negative energy transfer (with exactly the opposite behavior for the left side of Fig.~\ref{fig:fig2}).
Also visible
in Fig.~\ref{fig:fig2} is a less prominent small peak at about twice the energy of the BP and with a FWHM roughly twice as large.  This is interpreted as a double excitation of the BP feature and is denoted also by a dashed vertical red line.
The insert of Fig.~\ref{fig:fig2} shows the averages of all spectra taken over the region containing the BP and B2 features with the error bars representing one $\sigma$ standard deviation.  The BP is clearly visible as both a creation and annihilation event, and the double excitation B2 is also prominent on the energy gain (annihilation) side.  The B2 feature is less distinct with larger error bars on the energy loss (creation) side, but this is a well-known effect in He atom scattering; the $t^3$ Jacobian factor which must be applied to a TOF spectrum in order to transform it to an energy-resolved spcectrum greatly enhances the noise in the data as the energy loss increases.
It is of interest to note that the value we obtain for the BP, 6~\si{\milli \electronvolt}, is lower than values typically associated with the BP in bulk amorphous silica. One example is given in Ref.~\cite{Horbach}, where a molecular dynamics study predicts a BP frequency of 1.7~THz, roughly 7~\si{\milli \electronvolt}. This harmonises with the BP observation on the surface of vitreous silica at a lower frequency than in bulk~\cite{PhysRevLett.99.035503}. 

%----------------------------------------------------------------------------
% --- Figure
%----------------------------------------------------------------------------

 {
Figure~\ref{fig:fig3} shows  dispersion curves, i.e., plots of $\Delta E$ as a function of parallel wave vector transfer $\Delta Q$ for all inelastic features observed.  Figure~\ref{fig:fig3}(a) shows measurements for all polar angles taken at a single azimuthal angle of $12^\circ$, while Fig.~\ref{fig:fig3}(b) shows all features observed at all polar and azimuthal angles of the total 156 spectra data sets. Also shown in each panel of  Fig.~\ref{fig:fig3} are three representative scan curves shown as dashed green curves denoted by the corresponding incident polar angle. These curves are parabolas defined by the combined laws of conservation of energy and parallel momentum and their positions depend on the incident polar angle.  Shown in both panels of  Fig.~\ref{fig:fig3} are the scan curves for the largest and smallest angle measured, as well as the curve passing most closely to the specular position at $\theta_i = 45^\circ$. The importance of the scan curves is that for a given $\theta_i$ only those quantum features (inelastic or elastic) that the scan curves cross can be observed.
The blue diamond data points show clearly the strongly dispersive ZA mode previously measured~\cite{PhysRevLett.99.035503}.  The red circle data points give the positions of the BP mode, and it is clear that this is at very nearly constant energy for all $\Delta Q$, i.e., it is dispersionless as expected~\cite{PhysRevLett.99.035503}.
In the vicinity of the ZA mode the BP feature could not be distinguished because it is obscured by the much larger ZA peak.
Also clearly visible and denoted by inverted triangular data points is the dispersionless double excitation of the BP at about 12~meV. The larger scatter in the data for the double excitation is indicative of the fact that its FWHM is about twice as large as that of the BP.
}

%----------------------------------------------------------------------------
% --- Figure
%----------------------------------------------------------------------------
% ----------------------------------------------------------------------------

As mentioned above, it has been suggested that van Hove singularities could be the root cause of BP features in bulk materials, and in particular for the case of the observed BP in crystalline quartz~\cite{Chum-2011}. This does not seem to be the case for the present measurements.  The reason is that the van Hove singularities occur at very special points in the Brillouin zone, namely where the gradient with respect to wave vector of a phonon mode energy vanishes.  In the present case the phonon spectral density is measured as a function of wave vector $\Delta Q$.  As is evident in Figs.~\ref{fig:fig2} and~\ref{fig:fig3} the 2D BP feature is present, in roughly the same size and shape, over all measured values of $\Delta Q$.   This makes it unlikely that the present 2D BP is caused by a van Hove singularity, which should appear at only very special points in $\Delta Q$.  One could argue that backfolding of the Brillouin zone by some sort of super-periodicity could cause a van Hove singularity to appear to be spread out over a range of $\Delta Q$ values.  However, such backfolding would also affect the ZA mode, and this is clearly not happening as is apparent from Fig.~\ref{fig:fig3}.  

A further argument for not identifying the BP with a van Hove singularity is presented in Ref.~\cite{PhysRevB.98.174207}. In a 2D model glass they are able to differentiate between the BP and the van Hove singularity, showing that the BP in their system appears at a lower frequency. 

To conclude: we present the first measurements of the BP in a 2D material, 2D silica. We obtain a value of around 6~meV (1.5~THz), clearly different from 4~meV (1~THz), the value of the surface BP measured on bulk, vitreous silica~\cite{PhysRevLett.99.035503}. Thus our measurements constitute an example of a fundamental difference in the behavior of a  2D material compared to its bulk counterpart. Future work should include (i) measurements of bilayer silica grown on other substrates~\cite{huang2012} to access the influence of the substrate binding energy on the BP behaviour (ii) temperature dependent measurements to test if the blueshift behaviour observed for the surface  BP of bulk, vitreous silica is also found in 2D silica and (iii) measurements searching for the BP in other 2D (crystalline) materials. 
 
It is of serious interest to note that bilayer silica can be prepared in
different states of structural order~\cite{BUCHNER2017341}. The important implication is that bilayer silica is a material in which the BP can be studied as a function of crystalline order, ranging from completely ordered to disordered phases.

%Acknowledgements
\smallskip
\begin{acknowledgments}
It is a pleasure to thank Wolfram Steurer for useful discussions and
to acknowledge the valuable contributions and stimulating discussions with Gianfranco Pacchioni. 
H.-J. F. acknowledges funding from the European Research Council (ERC) under the European Union’s Horizon 2020 research and innovation program (Grant Agreement No. 669179).
B. H. acknowledges funding from the
Research Council of Norway (Projects No. 213453 and
No. 234159).
M. T. acknowledges funding from the
Research Council of Norway (Project No. 324183).
\end{acknowledgments}

% --------------------------------------------------------------------
% BIBLIOGRAPHY
% --------------------------------------------------------------------
%

 \end{document}